\documentclass[12pt]{article}

\usepackage{amsmath,amsfonts,amssymb}

\def\mH{\mathcal{H}}

\def\bx{\mathbf{x}}
\def\by{\mathbf{y}}
\def\bz{\mathbf{z}}

\newcommand{\hg}{\hat{g}}

\newcommand{\bT}{\mathbf{T}}

\def\pb #1{\left\{#1\right\}}

\begin{document}

\begin{titlepage}

\rightline{\footnotesize{}} \vspace{-0.2cm}

\begin{center}

\vskip 0.4 cm

\begin{center}
{\Large{ \bf Non-Local Gravity from  Hamiltonian
Point of View}}
\end{center}

\vskip 1cm

{\large Josef Kluso\v{n}$^{}$\footnote{E-mail:
{\tt klu@physics.muni.cz}}
}

\vskip 0.8cm

{\it
Department of
Theoretical Physics and Astrophysics\\
Faculty of Science, Masaryk University\\
Kotl\'{a}\v{r}sk\'{a} 2, 611 37, Brno\\
Czech Republic\\
[10mm]}

\vskip 0.8cm

\end{center}

\begin{abstract}
This short note is devoted to the canonical
analysis of the non-local theories of gravity. We
find their Hamiltonian and determine the
algebra of constraints. We perform this analysis
for non-local theories of gravity formulated
both in Jordan and Einstein frame. The result
of our analysis suggests that 
 Hamiltonian formulation does not bring to clear
identification of ghosts presence in non-local gravity.
\end{abstract}

\bigskip

\end{titlepage}

\newpage

\section{Introduction}\label{first}
Recent  experimental data suggests that
the expansion of the universe is
accelerating
\cite{Perlmutter:1998np,Komatsu:2010fb}.
One of the most popular approach how to
explain current expansion of the
universe is  the introduction of
cosmological constant dark energy in
the framework of general relativity
\footnote{For
review, see \cite{Weinberg:1988cp} and
the most recent \cite{Li:2011sd}.}. Another possibility how
to explain the acceleration of the universe
is to modify  of
gravity action. The most well known example
such a theory are  $F(R)$ theories of  gravity
where $R$ is the scalar curvature of
$D+1$ dimensional space-time and $F$ is
an arbitrary function, for review of
$F(R)$ gravity, see
\cite{Nojiri:2006ri,Nojiri:2010wj,DeFelice:2010aj,Faraoni}.

Another example of modifications of gravity
that could explain the current
acceleration \cite{Deser:2007jk}
are  non-local modifications of gravity.
This possibility is closely related to the
proposal presented
 in \cite{ArkaniHamed:2002fu} where
  authors suggested that
the cosmological constant problem could
be solved  in the context of non-local
gravity. This idea was further elaborated
in recent papers
\cite{Nojiri:2010pw,Bamba:2011ky}.
There are also additional
 reasons why it is interesting to study
the non-local modification of gravity. For
example,  non-local effective
field theories naturally emerge in the
framework of string field theory
\cite{Aref'eva:2004qr,Aref'eva:2007yr,Joukovskaya:2007nq,Calcagni:2007ef,Aref'eva:2011gj,
Galli:2010qx,Biswas:2010xq}
and hence the string theories could
provide natural UV completion of non-local
theories.
For  further analysis of non-local gravity from different
points of view,  see
\cite{Capozziello:2008gu,Cognola:2009jx,Nojiri:2007uq,Jhingan:2008ym}.

In summary,  non-local gravity models
are very intensive studied and deserve
to be investigated further
 from different point
of views. For example,  one would like to
see  how the non-local character of
given theory is reflected in its
canonical formulation.
 The goal of this paper is to
  perform  the Hamiltonian analysis of the
broad class of non-local theories of
gravity
\cite{Nojiri:2007uq}. We analyze these theories
in Jordan frame and then in  Einstein frame.
We determine the constraint structure of given theories
and we argue that they obey the standard rules
of geometrodynamics
\cite{Isham:1984sb,Isham:1984rz,Hojman:1976vp}
which is in agreement with the fact that these
theories are invariant under diffeomorphism
transformations.
On the other hand we show that the Hamiltonian
structure of given theories depends on the
character of the non-local action.
 More precisely,
due to the fact that these actions contain
derivative of scalar curvature
it is convenient to introduce the appropriate
number of scalar fields
\cite{Nojiri:2007uq} and rewrite these non-local
theories of gravity to the specific form of the
scalar tensor theories.
 Then the crucial point is whether the
 scalar field $A$  possesses
canonical conjugate momenta or not.
More precisely, for
 the action where $A$ appears linearly
but which is  general function of $\Box^{-1}A,\Box^{-2}A,\dots
$ we find that this theory possesses collection of
two second class constraints. The presence of
these constraints imply that the Poisson brackets
between canonical variables  should be replaced
with corresponding Dirac brackets. We also explicitly
show that these Dirac brackets depend on
phase space variables. This is very non-trivial
result whose origin can be traced to the
non-local character  of the theory. On the other hand
we show that the Dirac  algebra of the constraints takes again
the familiar form and obeys the standard rules of
geometrodynamics.

We should also stress one important point. The present
Hamiltonian analysis is not sensitive to the fact whether
some of the scalars have the kinetic term with negative
sign and hence should be considered as ghosts.
This is a consequence of the fact that the Hamiltonian
is linear combination of the first class constraints
that according to the basic principles of the theory
of constraints systems
\cite{Henneaux:1992ig}
should vanish on
the  constraint surface.

Let us outline our results. We perform the Hamiltonian
analysis of non-local theories of gravity and determine their
constraints structure. We find that the constraints
obey the standard rules of geometrodynamics. We also determine
the corresponding Dirac brackets between canonical
variables for particular form of the non-local gravity action.
We  derive equivalent results
when we consider theories formulated
 both in Jordan and in Einstein frame.

The structure of this paper is as follows. In the next
section (\ref{second})  we introduce
non-local theories of gravity and map them to their
 Jordan frame. Then we perform
their Hamiltonian analysis and discuss the canonical
structure of given theory with  dependence on the
properties of the scalar field $A$. In section (\ref{third})
we analyze these theories formulated in  Einstein frame and
find their canonical structure
and determine the results that are equivalent to the
ones derived in (\ref{second})
\footnote{We use units of $\hbar=c=1$ and denote
the gravitational constant $8\pi G$ by
$\kappa^2=\frac{8\pi}{M_{pl}^2}$ with
the Planck mass of $M_{pl}=G^{-1/2}=
1.2\times 10^{19}GeV$.}.

\section{Hamiltonian Analysis of
Non-Local Gravity}\label{second}
We begin with the action for  non-local gravity
that was recently studied in
\cite{Nojiri:2010pw,Bamba:2011ky}
\begin{equation}\label{nonlocal}
S=\int d^dx \sqrt{-\hg}
\left[\frac{1}{2\kappa^2}
{}^{(d)}R(\hg)(1+f(\Box^{-1}
{}^{(d)}R(\hg)))-2\Lambda\right]
\end{equation}
where $f$ is any function,${}^{(d)}R$ is
$d\equiv D+1$-dimensional scalar curvature,
  $\Box$ is
d'Alambertian
$\Box=\hg^{\mu\nu}\hat{\nabla}_\mu
\hat{\nabla}_\nu=\frac{1}{\sqrt{-\hg}}
\partial_\mu[\sqrt{-\hg}\hg^{\mu\nu}\partial_\nu]$,
 $\Box^{-1}$ is the
inverse of given operator and $\Lambda$
 is a cosmological constant
 \footnote{
 For simplicity we restrict ourselves to the analysis of
pure non-local theory
keeping in mind that it is
straightforward to generalize our analysis to
the case of when the matter contribution is present.}.
Due to the presence of the operator $\Box^{-1}$
it is convenient to introduce
 two scalar fields $\psi $ and $\xi$
 and rewrite the action
(\ref{nonlocal})
 in the following form
\begin{eqnarray}\label{nonlocali}
S
=\int d^{D+1} x \sqrt{-g}
\left[\frac{1}{2\kappa^2}
\left({}^{(d)}R(1+f(\psi)-\xi)-g^{\mu\nu}
\partial_\mu \xi\partial_\nu \psi
-2\Lambda\right)\right] \ .  \nonumber \\
\nonumber \\
\end{eqnarray}
It is easy task to show that
 the actions
 (\ref{nonlocali}) and
(\ref{nonlocal}) are equivalent. In fact,
the  variation of the action
(\ref{nonlocali}) with
respect to $\xi$ gives
\begin{equation}
\Box \psi={}^{(d)}R
\end{equation}
that implies $\psi=\Box^{-1}{}^{(d)}R$.
Then substituting this result into
(\ref{nonlocali}) we obtain
(\ref{nonlocal}).

In what follows we will be more general
and  consider
following general form of
non-local action \cite{Jhingan:2008ym}
\begin{eqnarray}\label{nonlocalgen}
S=\int d^{D+1}x \sqrt{-\hg} \left\{
\frac{1}{2\kappa^2} F({}^{(d)}
R,\Box {}^{(d)}R,\Box^2
{}^{(d)}R,\dots,\Box^m
{}^{(d)}R,\Box^{-1}{}^{(d)}R,
\Box^{-2}{}^{(d)}R,\dots,\Box^{-n}{}^{(d)}R)
\right\} \ .
\nonumber \\
\end{eqnarray}
As usual it is convenient to map given
action to more tractable form.
Following \cite{Jhingan:2008ym} we
firstly introduce scalar fields $A,B$
and rewrite the action
(\ref{nonlocalgen}) into the form
\begin{eqnarray}\label{nonlocalgen1}
S=\int d^{D+1}x \sqrt{-\hg} \left\{
\frac{1}{2\kappa^2} F(A,\Box A,\Box^2
A,\dots,\Box^m
A,\Box^{-1}A,\Box^{-2}A,\dots,\Box^{-n}A)+
B({}^{(d)}R-A)
\right\} \ .
\nonumber \\
\end{eqnarray}
As the next step we
define two  scalar
fields $\xi_1,\psi_1$
in order to eliminate $\Box^{-1}A$.  To do this we
add following term to the action
\begin{equation}
\int
d^{D+1}x \sqrt{-\hg}\xi_1 (A-\Box
\psi_1)=\int d^{D+1}x \sqrt{-\hg}
(\hg^{\mu\nu}\nabla_\mu \xi_1\nabla_\nu
\psi_1+\xi_1 A) \ .
\end{equation}
At the same time we
introduce two fields $\chi_1,\eta_1$
in order to eliminate $\Box A$ and
add following term to the action
\begin{equation}
\int d^{D+1}\sqrt{-\hg}\chi_1
(\eta_1-\Box A)= \int d^{D+1}
\sqrt{-\hg}(\hg^{\mu\nu} \partial_\mu
\chi_1\partial_\nu A+ \xi_1\eta_1) \
\end{equation}
so that  the action (\ref{nonlocalgen1})
takes the form
\begin{eqnarray}\label{nonlocalgen2}
S=\int d^{D+1}x \sqrt{-\hg} \left\{
\frac{1}{2\kappa^2} F(A,\eta_1,\Box
\eta_1,\dots,\Box^{m-1}\eta_1
,\psi_1,\Box^{-1}\psi,\dots,\Box^{-n+1}\psi)+
\right. \nonumber \\
+\left.B({}^{(d)}R-A)+ (\hg^{\mu\nu}
\partial_\mu \chi_1\partial_\nu A+
\chi_1\eta_1)+
(\hg^{\mu\nu}\partial_\mu
\xi_1\partial_\nu \psi_1+\xi_1 A)
\right\} \ .
\nonumber \\
\end{eqnarray}
From this analysis it is clear
 how to proceed further.
We introduce
following content of the scalar fields
$A,B,\chi_k,\eta_k, k=1,2,\dots,m$ and
$\xi_l,\psi_l,l=1,2,\dots,n$. Then repeating
the procedure  presented  above we
can rewrite the action (\ref{nonlocalgen})
into the form
\begin{eqnarray}\label{nonlocalgenfin}
S&=&\int d^{D+1}x \sqrt{-\hg} \left\{
\frac{1}{2\kappa^2}
F(A,\eta_1,\eta_2,\dots,\eta_m,\psi_1,
\psi_2,\dots,\psi_n)+
\right. \nonumber \\
&+&\left.B({}^{(d)}R-A)+ \hg^{\mu\nu}
\partial_\mu \chi_1\partial_\nu A+
\hg^{\mu\nu}\sum_{l=2}^m\partial_\mu
\chi_l\partial_\nu\eta_{l-1}+ \sum_{l=1}^m
\chi_l\eta_l+\right.\nonumber \\
& & \left.+ \hg^{\mu\nu}
 \sum_{l=1}^n\partial_\mu
\xi_l\partial_\nu \psi_l+\xi_1 A
+\sum_{l=2}^n \xi_l\psi_{l-1}
\right\} \ .
\nonumber \\
\end{eqnarray}
This form of the
action is our starting point
for the Hamiltonian
analysis of non-local theories of gravity.

As usual in order to formulate the
Hamiltonian analysis of theory coupled to gravity
we have to introduce $D+1$ formalism.
Explicitly, let us
  consider $D+1$ dimensional
manifold $\mathcal{M}$ with the
coordinates $x^\mu \ , \mu=0,\dots,D$
and where $x^\mu=(t,\bx) \ ,
\bx=(x^1,\dots,x^D)$. We presume that
this space-time is endowed with the
metric $\hat{g}_{\mu\nu}(x^\rho)$ with
signature $(-,+,\dots,+)$. Suppose that
$ \mathcal{M}$ can be foliated by a
family of space-like surfaces
$\Sigma_t$ defined by $t=x^0$. Let
$g_{ij}, i,j=1,\dots,D$ denotes the
metric on $\Sigma_t$ with inverse
$g^{ij}$ so that $g_{ij}g^{jk}=
\delta_i^k$. We further introduce the
operator $\nabla_i$ that is covariant
derivative defined with the metric
$g_{ij}$.
 We  introduce  the
future-pointing unit normal vector
$n^\mu$ to the surface $\Sigma_t$. In
ADM variables we have
$n^0=\sqrt{-\hat{g}^{00}},
n^i=-\hat{g}^{0i}/\sqrt{-\hat{g}^{
00}}$. We also define  the lapse
function $N=1/\sqrt{-\hat{g}^{00}}$ and
the shift function
$N^i=-\hat{g}^{0i}/\hat{g}^{00}$. In
terms of these variables we write the
components of the metric
$\hat{g}_{\mu\nu}$ as
\begin{eqnarray}
\hat{g}_{00}=-N^2+N_i g^{ij}N_j \ ,
\quad \hat{g}_{0i}=N_i \ , \quad
\hat{g}_{ij}=g_{ij} \ ,
\nonumber \\
\hat{g}^{00}=-\frac{1}{N^2} \ , \quad
\hat{g}^{0i}=\frac{N^i}{N^2} \ , \quad
\hat{g}^{ij}=g^{ij}-\frac{N^i N^j}{N^2}
\ .
\nonumber \\
\end{eqnarray}
Then it is easy to see that
\begin{equation}
\sqrt{-\det \hat{g}}=N\sqrt{\det g} \ .
\end{equation}
We further define the extrinsic
curvature
\begin{equation}
K_{ij}=\frac{1}{2N}
(\partial_t g_{ij}-\nabla_i N_j-
\nabla_j N_i) \ ,
\end{equation}
where $\nabla_i$ is the covariant derivative
calculated using the metric $g_{ij}$.
It is well known that the components of
the Riemann tensor can be written in
terms of ADM variables. For
example, in case of Riemann curvature
we have
\begin{equation}\label{R}
{}^{(d)}R=K^{ij}K_{ij}-K^2+R+\frac{2}{\sqrt{-\hat{g}}}
\partial_\mu(\sqrt{-\hat{g}}n^\mu K)-
\frac{2}{\sqrt{g}N}\partial_i
(\sqrt{g}g^{ij}\partial_j N) \ ,
\end{equation}
where $K=K_{ij}g^{ji}$ and where
$R$ is Riemann curvature
calculated using the metric $g_{ij}$.
Note that $n^\mu$ has components
\begin{equation}
n^0=\frac{1}{N} \ , n^i=-\frac{N^i}{N} \ .
\end{equation}
Implementing $D+1$ formalism in the
 action (\ref{nonlocalgenfin}) we find that
 it has the form
\begin{eqnarray}\label{nonlocalgenfinD1}
S&=&\int d^{D+1}x N \sqrt{g}\left\{
\frac{1}{2\kappa^2}
F(A,\eta_1,\eta_2,\dots,\eta_m,\psi_1,
\psi_2,\dots,\psi_n)+\right.
 \nonumber \\
&+&B(K^{ij}K_{ij}-K^2+R-A)
-2\nabla_n B K-\frac{2}{\sqrt{g}}
\partial_j
(\sqrt{g}g^{ij}
\partial_j B)- \nonumber \\
&-& \nabla_n \chi_1\nabla_n A+g^{ij}
\partial_i \chi_1\partial_j A+
\sum_{l=2}^m (-\nabla_n \chi_1
\nabla_n\eta_{l-1}+ g^{ij}
\partial_i\chi_l\partial_j \eta_{l-1})
+ \sum_{l=1}^m
\chi_l\eta_l+\nonumber \\
&+&\left.
 \sum_{l=1}^n
(-\nabla_n \xi_l \nabla_n \psi_l
+g^{ij}
\partial_i \xi_l\partial_j\psi_l)
+\xi_1 A +\sum_{l=2}^n \xi_l\psi_{l-1}\right\} \ .
\nonumber \\
\end{eqnarray}
Using the form of the action
 (\ref{nonlocalgenfinD1}) we can proceed
 to the Hamiltonian formalism. Explicitly,
 from (\ref{nonlocalgenfinD1})
 we determine  conjugate momenta
\begin{eqnarray}
\pi_N &\approx & 0 \ , \quad \pi_i\approx 0
\ , \quad  \pi^{ij}= B
\sqrt{g}(K^{ij}-g^{ij}K)
-\sqrt{g}\nabla_n Bg^{ij} \ , \quad
\nonumber \\
 p_B &=&-2\sqrt{g}K \ , \quad
p_A=-\sqrt{g}\nabla_n\chi_1 \ ,
\nonumber \\
p_{\chi_l}&=&-\sqrt{g}\nabla_n\eta_{l-1}
\ ,  \quad  p_{\eta_{l-1}}=
-\sqrt{g}\nabla_n \chi_l \ ,
l=2,\dots,m \ ,
\nonumber \\
p_{\xi_k}&=&-\sqrt{g}\nabla_n\psi_k \ , \quad
p_{\psi_k}=-\sqrt{g}\nabla_n\xi_k \ ,
k=1,\dots,n  \ . \nonumber \\
\end{eqnarray}
Then after some algebra we find the
 Hamiltonian in the form
\begin{equation}
H=\int d^D\bx (N\mH_T+N^i\mH_i) \ ,
\end{equation}
where
\begin{eqnarray}
\mH_T&=&
\frac{1}{\sqrt{g}B}\pi^{ij}g_{ik}g_{il}\pi^{kl}
-\frac{1}{\sqrt{g}BD}\pi^2-\frac{\pi
p_B}{\sqrt{g}D}
+\nonumber \\
&+&\frac{B}{4\sqrt{g}D}(D-1)p_B^2-\sqrt{g}BR+
2\partial_i[\sqrt{g}g^{ij}\partial_j
B]-
\nonumber \\
&-&\frac{1}{\sqrt{g}} p_A p_{\chi_1}
-\frac{1}{\sqrt{g}}\sum_{l=2}^m
p_{\chi_l}p_{\eta_{l-1}}-\frac{1}{
\sqrt{g}}\sum_{k=1}^n
p_{\xi_k}p_{\psi_k}
-\nonumber \\
 &-& \sqrt{g}
\frac{1}{2\kappa^2}
F(A,\eta_1,\eta_2,\dots,\eta_m,\psi_1,
\psi_2,\dots,\psi_n)+\sqrt{g}BA
+2\partial_j [\sqrt{g}g^{ij}
\partial_j B]- \nonumber \\
&-&\sqrt{g}g^{ij}
\partial_i \chi_1\partial_j A-
\sqrt{g}g^{ij} \sum_{l=2}^m
\partial_i\chi_l\partial_j \eta_{l-1}
-\sqrt{g} \sum_{l=1}^m
\chi_l\eta_l-\nonumber \\
&-&
\sqrt{g}g^{ij} \sum_{l=1}^n
\partial_i \xi_l\partial_j\psi_l
-\sqrt{g}\xi_1 A -\sqrt{g}\sum_{l=2}^n \xi_l\psi_{l-1} \ ,
\nonumber \\
 \mH_i&=&-2g_{ik}\nabla_l\pi^{kl}+
 p_A\partial_i A+
 p_B\partial_i B+
 \sum_{l=1}^m
p_{\chi_l}\partial_i \chi_l
+\sum_{l=2}^m
p_{\eta_l}\partial_i\eta_l +
\sum_{k=1}^n(
 p_{\xi_k}\partial_i
\xi_k+p_{\psi_k}\partial_i\psi_k) \ ,
\nonumber \\
\end{eqnarray}
and where $\pi=\pi^{ij}g_{ji}$.
As usual the requirement of the preservation of
the primary constraints $\pi_N\approx 0,
\pi_i\approx 0$ implies the secondary one
\begin{equation}
\mH_T\approx 0 \ , \quad \mH_i\approx 0 \ .
\end{equation}
As the next step we have to check the consistency
of the secondary constraints with the time
development of the system. For that reason
it is convenient to
introduce the smeared form of these
constraints
\begin{equation}\label{defbT}
\bT_T(N)=\int d^D\bx N(\bx)\mH_T(\bx) \
, \quad \bT_S(N^i)=\int d^D\bx
N^i(\bx)\mH_i(\bx) \ .
\end{equation}
 Then using the canonical
Poisson brackets
\begin{eqnarray}
\pb{g_{ij}(\bx),\pi^{kl}(\by)}&=&
\frac{1}{2}(\delta_i^k\delta_j^l+
\delta_i^l\delta_j^k)\delta(\bx-\by) \
, \nonumber \\
\pb{A(\bx),p_A(\by)}&=& \delta(\bx-\by) \
, \quad
\pb{B(\bx),p_B(\by)}=\delta(\bx-\by) \
, \nonumber \\
\pb{\chi_l(\bx),p_{\chi_k}(\by)}&=&
\delta_{lk}\delta(\bx-\by) \ , \quad \pb{
\eta_l(\bx),p_{\eta_k}(\by)}=
\delta_{lk}\delta(\bx-\by) \ ,
\nonumber \\
\pb{\xi_l(\bx),p_{\xi_k}(\by)}&=&
\delta_{lk}\delta(\bx-\by) \ , \quad
\pb{\psi_l(\bx),p_{\psi_k}(\by)}=
\delta_{lk}\delta(\bx-\by) \
\nonumber \\
\end{eqnarray}
we easily determine the well known  algebra of
constraints \cite{Isham:1984sb,Isham:1984rz,Hojman:1976vp}
\begin{eqnarray}\label{pbbTTS}
\pb{\bT_T(N),\bT_T(M)}&=&
\bT_S(N\partial^i M-M\partial^i N) \ ,
\nonumber \\
\pb{\bT_S(M^i),\bT_T(N)}&=&
\bT_T(M^i\partial_i N ) \ , \nonumber
\\
\pb{\bT_S(M^i),\bT_S(N^j)}&=&
\bT_S(N^i\partial_i M^j-M^i\partial_i
N^j) \ . \nonumber \\
\end{eqnarray}
In other words the constraints
(\ref{defbT}) are preserved during
the time evolution of the system. Note
also that these constrains have to vanish
weakly. As a result
the Hamiltonian has to vanish on the constraint
surface and hence any instability related
to the presence of the ghosts (which is general
property of any non-local theory) is
not seen on the level of classical Hamiltonian
analysis.

It is important to stress that
during the analysis performed above we implicitly
presumed that there is a momentum conjugate
to $A$. However it turns out that for the non-local
actions that do not depend on $\Box {}^{(d)}R$ the momentum
conjugate to $A$ is absent. More precisely, let us consider the action
\begin{equation}\label{nonlocatr}
S=\frac{1}{2\kappa^2}
\int d^{D+1}x\sqrt{-\hg}F({}^{(d)}R,\Box^{-1}{}^{(d)} R,
\Box^{-2}
{}^{(d)}R,\dots,
\Box^{-n} {}^{(d)}R)
\end{equation}
that is the generalization of the
 action (\ref{nonlocali}).
 Performing the same
 analysis as above we find that this action
 takes the form
\begin{eqnarray}\label{nonlocalgenfinDn}
S&=&\int d^{D+1}x   \sqrt{g}N\left\{
\frac{1}{2\kappa^2}
F(A ,\psi_1,
\psi_2,\dots,\psi_n)+\right.
 \nonumber \\
&+& \left. B(K^{ij}K_{ij}-K^2+R-A)
-2\nabla_n B K-\frac{2}{\sqrt{g}}\partial_j
(\sqrt{g}g^{ij}
\partial_j B)+\right. \nonumber \\
&+& \left.  \sum_{l=1}^n
(-\nabla_n \xi_l \nabla_n \psi_l
+g^{ij}
\partial_i \xi_l\partial_j\psi_l)
+\xi_1 A +\sum_{l=2}^n \xi_l\psi_{l-1}
\right\} \ .
\nonumber \\
\end{eqnarray}
From
the action (\ref{nonlocalgenfinDn}) we find the
conjugate momenta
\begin{eqnarray}
\pi_N &\approx& 0 \ , \quad \pi^i\approx 0
\ , \quad \pi^{ij}= B
\sqrt{g}(K^{ij}-g^{ij}K)
-\sqrt{g}\nabla_n Bg^{ij} \ ,
\nonumber \\
 p_B&=&-2\sqrt{g}K \ , \quad
p_A\approx 0 \ ,
\nonumber \\
p_{\xi_k}&=&-\sqrt{g}\nabla_n\psi_k \ , \quad
p_{\psi_k}=-\sqrt{g}\nabla_n\xi_k \ ,
l=1,\dots,n \nonumber \\
\end{eqnarray}
and the Hamiltonian
\begin{equation}
H=\int d^D\bx (N\mH_T+N^i\mH_i+v^Ap_A) \ ,
\end{equation}
where
\begin{eqnarray}
\mH_T&=&
\frac{1}{\sqrt{g}B}\pi^{ij}g_{ik}g_{il}\pi^{kl}
-\frac{1}{\sqrt{g}BD}\pi^2-\frac{\pi
p_B}{\sqrt{g}D}
+\nonumber \\
&+&\frac{B}{4\sqrt{g}D}(D-1)p_B^2-\sqrt{g}BR+
2\partial_i[\sqrt{g}g^{ij}\partial_j
B]-\frac{1}{
\sqrt{g}}\sum_{k=1}^n
p_{\xi_k}p_{\psi_k}
-\nonumber \\
&-& \sqrt{g}
\frac{1}{2\kappa^2}
F(A,\psi_1,
\psi_2,\dots,\psi_n)+\sqrt{g}BA
- \nonumber \\
&-&\sqrt{g} \sum_{l=1}^n
g^{ij}
\partial_i \xi_l\partial_j\psi_l
-\sqrt{g}\xi_1 A -\sqrt{g}\sum_{l=2}^n \xi_l\psi_{l-1} \ ,
\nonumber \\
 \mH_i&=&-2g_{ik}\nabla_l\pi^{kl}+
 p_A\partial_i A+
 p_B\partial_i B
 +
\sum_{k=1}^n(
 p_{\xi_k}\partial_i
\xi_k+p_{\psi_k}\partial_i\psi_k) \ .
\nonumber \\
\end{eqnarray}
We again  introduce the
smeared constraints $\bT_T(N),\bT_S(N^i)$ and
we easily find that they obey the relations
(\ref{pbbTTS}).
 On the other
hand the requirement of the preservation
of the  primary constraint $p_A\approx 0$
implies the secondary one:
\begin{eqnarray}
\partial_t p_A=\pb{p_A,H}\approx
N\sqrt{g}\left(\frac{1}{2\kappa^2}
\frac{dF}{dA}-B+\xi_1\right)\equiv N G_A\approx 0 \ .
\nonumber \\
\end{eqnarray}
Finally we determine time evolution  of
the constraint $G_A$
\begin{eqnarray}\label{eqvA1}
\partial_t G_A&=&\pb{G_A,H}=
\nonumber \\
&=&
N\left(-\frac{1}{2\kappa^2}
\sum_{l=1}^n\frac{d^2 F}{dA d\psi_l}
p_{\xi_l}-p_{\psi_1}+\frac{\pi}{D}
-\frac{B}{2D}(D-1)p_B\right)+
\frac{1}{2\kappa^2}
\frac{d^2 F}{dA^2}v^A=0 \ .
\nonumber \\
\end{eqnarray}
We observe  that there are two possible
alternatives. The first one corresponds
to the situation
when  $\frac{d^2F}{d^2A}\neq 0$
and we see that the equation
(\ref{eqvA}) uniquely fixes the value of the
Lagrange multiplier $v^A$. Then  we can
finish the analysis of the
 consistency of constraints with the time
evolution of the system since now
$p_A\approx 0 \ , G_A\approx 0$ are the second class
constraints with non-zero Poisson bracket
\begin{equation}
\pb{p_A(\bx),G_A(\by)}=
-\frac{1}{2\kappa^2}
\frac{d^2F}{d^2A}(\bx)\delta(\bx-\by) \ .
\end{equation}
 In principle
these constraints can be solved for $p_A$ and
$A$ and hence we find theory that has the same
physical
content as the $F(R)$ theory of gravity coupled with the
collection of the scalar fields.
It is also easy to see that
 the Dirac brackets of  canonical
variables that define reduced phase space
 coincide with the Poisson brackets.

The more interesting example  corresponds to the second
 situation when
$\frac{d^2F}{d^2A}=0$ so that $F$ has linear dependence on
$A$
\begin{equation}
F=AU_0(\psi_1,\dots,\psi_n)+U_1(\psi_1,\dots,\psi_n) \
\end{equation}
and hence the constraint $G_A$ has explicit form
\begin{equation}\label{GA}
G_A=\sqrt{g}\left(\frac{1}{2\kappa^2}
U_0-B+\xi_1\right)\approx 0 \ .
\end{equation}
Then the equation (\ref{eqvA1})
implies an additional  constraint
\begin{equation}
G_A^{II}=
-\frac{1}{2\kappa^2}
\sum_{l=1}^n\frac{dU_0}{ d\psi_l}
p_{\xi_l}-p_{\psi_1}+\frac{\pi}{D}
-\frac{B}{2D}(D-1)p_B\approx 0 \ .
\end{equation}
Note that the Poisson bracket
between $G_A$ and $G_A^{II}$ is
equal to
\begin{eqnarray}\label{deftriangle}
\pb{G_A(\bx),G_A^{II}(\by)}=
\sqrt{g}\left(-\frac{1}{\kappa^2}
\frac{dU_0}{d\psi_1}+
\frac{D-1}{D}B\right)\delta(\bx-\by)\equiv
\triangle(\bx)\delta(\bx-\by) \ .
\nonumber \\
\end{eqnarray}
The next step is to explicitly solve
the constraints $p_A\approx 0$ and
$G_A\approx 0,\ G_A^{II}\approx 0$.
Since the constraint $p_A\approx 0$ is
the first class constraint we impose
the gauge fixing condition $A=const$.
As a result the pair $A,p_A$ is eliminated
from the theory.
  On the
other hand we solve the constraint
$G_A$ for $B$ and we find
\begin{equation}\label{B}
B=\frac{1}{2\kappa^2}U_0+\xi_1 \
.
\end{equation}
In the same way we solve the constraint
$G_A^{II}$ for $p_B$ with the result
\begin{eqnarray}\label{pB}
p_B
=\frac{2D}{(D-1)}
\frac{1}{
\frac{1}{2\kappa^2}U_0+\xi_1}
\left(\frac{\pi}{D}
-\frac{1}{2\kappa^2}
\sum_{l=1}^n\frac{dU_0}{ d\psi_l}
p_{\xi_l}+p_{\psi_1}\right) \ .
\nonumber \\
\end{eqnarray}
As the final point we have to  replace the
Poisson brackets with corresponding Dirac brackets.
However it is  important to stress
that the Dirac brackets between the first
class constraints coincide with corresponding
Poisson brackets. Let us demonstrate
this claim on the  following  example
\begin{eqnarray}
\pb{\bT_T(N),\bT_T(M)}_D&=&
\pb{\bT_T(N),\bT_T(M)}-\nonumber \\
&-&
\int d^D\bz d^D\bz'
\pb{\bT_T(N),G_A(\bz)}
\triangle^{-1}(\bz,\bz')\pb{G_A^{II}(\bz'),
\bT_T(M)}+\nonumber \\
&+& \int d^D\bz d^D\bz'
\pb{\bT_T(N),G^{II}_A(\bz)}
\triangle^{-1}(\bz,\bz')\pb{G_A(\bz'),
\bT_T(M)}\approx \nonumber \\
& \approx &
\pb{\bT_T(N),\bT_T(M)}
\nonumber \\
\end{eqnarray}
due to the fact that $\pb{\bT_T(N),G_A(\bz)}
=-N(\bz)G_A^{II}(\bz)\approx 0$. In the same way we can show
that the  Dirac brackets $\pb{\bT_T(N),\bT_S(N^i)}_D,
\pb{\bT_S(N^i),\bT_S(M^j)}_D$  coincide with corresponding
Poisson brackets. Further, it is also
 easy to see that the Dirac brackets
between $g_{ij},\pi^{kl}$ coincides
with the Poisson brackets again simply
from the fact that $\pb{g_{ij},G_A}=0 \ ,
\pb{\pi^{ij},G_A}\approx 0$.
 On the other
hand the situation is more complicated
in case of the modes
$\xi_l,\psi_l$ and corresponding
conjugate momenta $p_{\xi_l},p_{\psi_l}$.
Explicitly
\begin{eqnarray}
\pb{\xi_l(\bx),p_{\xi_k}(\by)}_D&=&
\pb{\xi_l(\bx),p_{\xi_k}(\by)}- \int d^D\bz
d^D\bz' \pb{\xi_l(\bx),G_A(\bz)}
\triangle^{-1}(\bz,\bz')
\pb{G_A^{II}(\bz'),p_{\xi_k}(\by)}+
\nonumber \\
&+&\int d^D\bz d^D\bz'
\pb{\xi_l(\bx),G_A^{II}(\bz)}
\triangle^{-1}(\bz,\bz')
\pb{G_A(\bz'),p_{\xi_k}(\by)}=
\nonumber \\
&=& \delta(\bx-\by)\delta_{lk}-\frac{1}{2\kappa^2}
\frac{dU_0}{d\psi_l}\triangle^{-1}
\sqrt{g}\delta_{1,k}\delta(\bx-\by) \ ,
\nonumber \\
\end{eqnarray}
where $\triangle^{-1}$ is defined
by the  equation
\begin{equation}
\int d^D\bz \triangle(\bx,\bz)
\triangle^{-1}(\bz,\by)=
\delta(\bx-\by) \ .
\end{equation}
Then using (\ref{deftriangle})
we find
\begin{equation}
\triangle^{-1}(\bx,\by)=
\frac{1}{\sqrt{g}(-\frac{1}{\kappa^2}
\frac{dU_0}{d\psi_1}+
\frac{D-1}{D}B)}
\delta(\bx-\by) \ .
\end{equation}
In the same way we find
\begin{eqnarray}
\pb{\xi_l(\bx),p_{\psi_k}(\by)}_D&=&
-\frac{\sqrt{g}}{4\kappa^4}
\frac{dU_0}{d\psi_l}\frac{dU_0}{d\psi_k}
\sqrt{g}\triangle^{-1}\delta(\bx-\by) \ ,
\nonumber \\
\pb{\psi_l(\bx),p_{\xi_k}(\by)}_D&=&
-\sqrt{g}\delta_{l,1}\delta_{k,1}\triangle^{-1}
\delta(\bx-\by) \ ,
 \nonumber \\
\pb{\psi_l(\bx),p_{\psi_k}(\by)}_D&=&
\delta(\bx-\by)\delta_{lk}-\frac{\sqrt{g}}{2\kappa^2}
\delta_{l,1}\frac{dU_0}{d\psi_k}\triangle^{-1}
\delta(\bx-\by) \ .
\nonumber \\
\end{eqnarray}
Remarkably the
presence of the second class constraints
implies  non-trivial Dirac brackets between
metric variables and scalar fields and corresponding
conjugate momenta. For example,
\begin{eqnarray}
\pb{g_{ij}(\bx),p_{\xi_l}(\by)}&=&
\int d^D\bz d^D\bz'
\pb{g_{ij}(\bx),G_A^{II}(\bz)}
\triangle^{-1}(\bz,\bz')
\pb{G_A(\bz'),p_{\xi_l}(\by)}=
\nonumber \\
&=&\frac{1}{D}g_{ij}\triangle^{-1}
\delta_{1,l}\delta(\bx-\by) \ .
\nonumber \\
\end{eqnarray}
In the same way we find
\begin{eqnarray}
\pb{g_{ij}(\bx),p_{\psi_k}(\by)}_D&=&
\frac{1}{2\kappa^2 D}g_{ij}
\frac{dU_0}{d\psi_k}\triangle^{-1}\delta(\bx-\by) \  ,
\nonumber \\
\pb{\pi^{ij}(\bx),p_{\xi_l}(\by)}_D&=&
-\frac{1}{D}\pi^{ij}\triangle^{-1}
\delta_{1,l}\delta(\bx-\by) \ ,
\nonumber \\
\pb{\pi^{ij}(\bx),p_{\psi_k}(\by)}_D&=&
-\frac{1}{2\kappa^2 D}\pi^{ij}
\frac{dU_0}{d\psi_k}\triangle^{-1}\delta(\bx-\by) \ .
\nonumber \\
\end{eqnarray}
Let us outline the results derived
in this section. We performed the
canonical formalism for non-local theories of
gravity. We found that the Hamiltonian
is given as a linear combination
of the first class constraints
with standard Poisson brackets. On the
other hand the Hamiltonian constraint
and symplectic structure defined on
the reduced phase space are very complicated
due to the relations (\ref{B}), (\ref{pB}).

In the next section we perform the
Hamiltonian analysis of non-local
theory of gravity that is formulated
in  the Einstein
frame.
\section{Non-local Gravity in Einstein
Frame}\label{third}
For some purposes it is convenient to
transform non-local theory to the
Einstein frame formulation
\footnote{For review, see for example
 \cite{Nojiri:2010pw,Bamba:2011ky}.}.
 Recall that under the scaling transformation
of metric
\begin{equation}\label{Weylg}
\bar{\hg}_{\mu\nu}=\Omega^2
\hg_{\mu\nu} \
\end{equation}
the scalar curvature transforms as
\begin{equation}
{}^{(d)}
\bar{R}=\frac{1}{\Omega^2}
\left({}^{(d)}R
-2D\frac{1}{\Omega}
\hg^{\mu\nu}\nabla_\mu \nabla_\nu
\Omega +D(3-D)\frac{\nabla_\mu
\Omega \nabla_\nu \Omega
\hg^{\mu\nu}}{\Omega^2}\right) \ .
\end{equation}
Let us  consider the most general
form of the non-local action
(\ref{nonlocalgenfin}). Then using
(\ref{Weylg})
with  $\Omega=B^{\frac{1}{1-D}}$ we can
 map the action
(\ref{nonlocalgenfin}) into the form
\begin{eqnarray}\label{nonlocalgenfinEF}
S&=&\int d^{D+1}x \sqrt{-\hg} \left\{
\frac{1}{2\kappa^2}
F(A,\eta_1,\eta_2,\dots,\eta_m,\psi_1,
\psi_2,\dots,\psi_n)+
\right. \nonumber \\
&+&\left. {}^{(d)} R
+\frac{1+D}{1-D}\frac{1}{B^2}\hg^{\mu\nu}
\partial_\mu B\partial_\nu B
-B^{\frac{2}{1-D}}
A+\right. \nonumber \\
&+& \left. \frac{1}{B}\hg^{\mu\nu}
\partial_\mu \chi_1\partial_\nu A+
\frac{1}{B}\hg^{\mu\nu}\sum_{l=2}^m\partial_\mu
\chi_l\eta_{l-1}+ B^{\frac{D+1}{1-D}}\sum_{l=1}^m
\chi_l\eta_l+\right.\nonumber \\
&+&\left. \frac{1}{B}\hg^{\mu\nu}
 \sum_{l=1}^n\partial_\mu
\xi_l\partial_\nu \psi_l+B^{\frac{D+1}{1-D}}\xi_1 A
+B^{\frac{1+D}{1-D}}\sum_{l=2}^n \xi_l\psi_{l-1}\right\} \ .
\nonumber \\
\end{eqnarray}
This is the non-local gravity action formulated in the
Einstein frame. Our goal is to perform
the Hamiltonian analysis of given action.

The simplest possibility  corresponds to the
situation when
 $\partial_\mu A\neq 0$.  In this case
 the
action  (\ref{nonlocalgenfinEF})
 has the structure
\begin{eqnarray}
S=\frac{1}{2\kappa^2} \int d^D x
\sqrt{-g} [ {}^{(d)}R
-\hg^{\mu\nu}G_{AB}(\Phi)\partial_\mu\Phi^A
\partial_\nu \Phi^B-V(\Phi)] \ ,  \nonumber \\
\end{eqnarray}
where $G_{AB}$ is a  specific field
dependent metric on the field space.
This is well known form of the scalar tensor
theory and it is simple task to
determine corresponding Hamiltonian
\begin{eqnarray}
H&=&\int d^D\bx (N\mH_T+N^ i\mH_i) \ ,
\quad \mH_T=\mH_T^{G.R.}+\mH_T^{scal} \
, \nonumber \\
\mH_T^{G.R.}
&=&\frac{4\kappa^2}{\sqrt{g}}
\left(\pi^{ij}\pi_{ij}+\frac{1}{1-D}\pi^2\right)
-\frac{1}{2\kappa^2}\sqrt{g}R \ ,
\nonumber \\
 \mH^{scal}_T&=& \frac{1}{2}
\left(\frac{\kappa^2}{\sqrt{g}} p_A
G^{AB}p_B+\frac{\sqrt{g}}{\kappa^2}
G_{AB}g^{ij}\partial_i\Phi^A\partial_j\Phi^B+V(\Phi)\right) \ ,
\nonumber \\
\mH_i&=& p_A\partial_i\Phi^A-2g_{ik}\nabla_j\pi^{jk}
\ . \nonumber
\\
\end{eqnarray}
Then  the standard analysis
implies that $\mH_T$ and $\mH_i$ are
the first class constraints and their
algebra takes the  form
(\ref{pbbTTS}). Recall again that the
Hamiltonian vanishes on the constraint
surface.

More interesting situation occurs in
case when  $\partial_\mu A=0$ which corresponds
to the  form of the non-local
gravity action (\ref{nonlocatr}).
 Following standard
analysis we derive the  Hamiltonian
in the form
\begin{equation}
H=\int d^D\bx (N\mH_T+N^i\mH_i) \ ,
\end{equation}
where
\begin{eqnarray}
\mH_T&=&
\frac{4\kappa^2}{\sqrt{g}}
\left(\pi^{ij}\pi_{ij}+\frac{1}{1-D}\pi^2\right)
-\frac{1}{2\kappa^2}\sqrt{g}R
-\nonumber \\
 &-& \sqrt{g}
\frac{1}{2\kappa^2}B^{\frac{1+D}{1-D}}
F(A,\psi_1,
\psi_2,\dots,\psi_n)+\sqrt{g}B^{\frac{2}{1-D}}A
-\frac{1-D}{1+D}\frac{B^2}{4\sqrt{g}}p_B^2
+\nonumber \\
&+& \frac{1+D}{1-D}\sqrt{g}\frac{1}{B^2}
g^{ij}\partial_i B\partial_j B
-\frac{
B}{
\sqrt{g}}\sum_{k=1}^n
p_{\xi_k}p_{\psi_k}-
\nonumber \\
&-&\frac{\sqrt{g}}{B}g^{ij}
 \sum_{l=1}^n\partial_i
\xi_l\partial_j\psi_l -\sqrt{g}
B^{\frac{D+1}{1-D}}\xi_1 A
-\sqrt{g}B^{\frac{1+D}{1-D}}\sum_{l=2}^n \xi_l\psi_{l-1} \ .
\nonumber \\
\end{eqnarray}
As usual we obtain the secondary
constraints $\mH_T,\mH_i$  that obey
the relations (\ref{pbbTTS}).
 On the other
hand the requirement of the preservation
of the  primary constraint $p_A\approx 0$
implies  the secondary one:
\begin{eqnarray}
\partial_t p_A=\pb{p_A,H}\approx
NB^{\frac{1+D}{1-D}}\sqrt{g}\left(\frac{1}{2\kappa^2}
\frac{dF}{dA}-B+\xi_1\right)\equiv N B^{\frac{1+D}{1-D}}
G_A\approx 0 \  ,
\nonumber \\
\end{eqnarray}
where $G_A$ coincides with the constraint
(\ref{GA}).
Finally we determine the time evolution  of
the constraint $G_A$
\begin{eqnarray}\label{eqvA}
\partial_t G_A&=&\pb{G_A,H}=\nonumber \\
&=&BN\left(-\frac{1}{2\kappa^2}
\sum_{l=1}^n\frac{d^2 F}{dA d\psi_l}
p_{\xi_l}- p_{\psi_1}
+\frac{1-D}{2(1+D)}B p_B\right)+
\frac{1}{2\kappa^2}
\frac{d^2 F}{dA^2}v^A=0 \ .
\nonumber \\
\end{eqnarray}
We observe that this equation possesses two
possible alternatives exactly as the
equation (\ref{eqvA1}).Since the analysis is completely
the same as the analysis presented below this equation
 we will not repeat it.

In summary, Einstein or Jordan frame formulation
of non-local theory of gravity leads to well defined
Hamiltonian systems where the Hamiltonian
is given as a linear combination of the constraints.
Due to the fact that these constraints have
to vanish weakly it does not matter whether the
scalars are ghosts or ordinary scalar fields
at least on the classical level.
\vskip 5mm

 \noindent {\bf
Acknowledgements:}
I would like to thank S. Odintsov for valuable
discussions during the my work in this project.
 This work   was also
supported by the Czech Ministry of
Education under Contract No. MSM
0021622409.
 \vskip 5mm

\end{document}